# Experimental verification of stopping-power prediction from single- and dual-energy computed tomography in biological tissues

**Christian Möhler[1,2], Tom Russ[1,2], Patrick Wohlfahrt[3,4], Alina Elter[1,2], Armin Runz[1,2], Christian Richter[3,4] and Steffen Greilich[1,2]**

[1] German Cancer Research Center (DKFZ), Heidelberg, Germany
[2] National Center for Radiation Research in Oncology (NCRO), Heidelberg Institute for Radiation Oncology (HIRO)
[3] OncoRay - National Center for Radiation Research in Oncology, Faculty of Medicine and University Hospital Carl Gustav Carus, Technische Universität Dresden Helmholtz-Zentrum Dresden - Rossendorf, Dresden, Germany
[4] Helmholtz-Zentrum Dresden - Rossendorf, Institute of Radiooncology - OncoRay, Dresden, Germany

E-mail: c.moehler@dkfz-heidelberg.de

**Abstract.** An experimental setup for consecutive measurement of ion and x-ray absorption in tissue or other materials is introduced. With this setup using a 3D-printed sample container, the reference stopping-power ratio (SPR) of materials can be measured with an uncertainty of below 0.1%. A total of 65 porcine and bovine tissue samples were prepared for measurement, comprising five samples each of 13 tissue types representing about 80% of the total body mass (three different muscle and fatty tissues, liver, kidney, brain, heart, blood, lung and bone). Using a standard stoichiometric calibration for single-energy CT (SECT) as well as a state-of-the-art dual-energy CT (DECT) approach, SPR was predicted for all tissues and then compared to the measured reference. With the SECT approach, the SPRs of all tissues were predicted with a mean error of $(-0.84 \pm 0.12)\%$ and a mean absolute error of $(1.27 \pm 0.12)\%$. In contrast, the DECT-based SPR predictions were overall consistent with the measured reference with a mean error of $(-0.02 \pm 0.15)\%$ and a mean absolute error of $(0.10 \pm 0.15)\%$. Thus, in this study, the potential of DECT to decrease range uncertainty could be confirmed in biological tissue.

# 1 Introduction

Treatment planning for radiotherapy with protons or heavier ions requires an accurate 3D map of particle stopping-power ratios (SPRs) of the patient. Nowadays, this map is clinically derived by x-ray computed tomography (CT), applying a piecewise linear transformation of CT numbers to SPR, often referred to as "Hounsfield look-up table" (HLUT). In practice, a HLUT is obtained empirically by fitting or interpolating the CT and SPR values of selected real or virtual calibration materials, which can either be measured (Jäkel *et al* 2001) or calculated from elemental composition in the so-called "stoichiometric calibration" (Schneider *et al* 1996).

As a promising alternative modality for particle treatment planning, dual-energy CT (DECT) is currently being investigated and clinically implemented (van Elmpt *et al* 2016, Wohlfahrt *et al* 2017b). Exploiting the energy dependence of photon absorption in matter using two different energy spectra enables the determination of electron density relative to water (Rutherford *et al* 1976, Brooks 1977). According to the Bethe formula (ICRU 1993), electron density enters linearly in SPR and dominates its variability for human tissue. As the second factor in the Bethe formula, the so-called stopping number, which contains the dependency on the mean excitation energy ("I-value"), has no direct analogue in keV photon absorption and thus requires an empirical proxy, such as the effective atomic number (Yang *et al* 2010) or relative photon absorption cross section (Möhler *et al* 2016). This implies that SPR prediction with DECT still contains an empirical component, even though its impact on prediction accuracy is strongly mitigated compared to SECT.

Consequently, the uncertainty in both SECT- and DECT-based SPR prediction depends on empirical knowledge of radiological properties of human tissue. A validation and comparison of SPR prediction methods in real tissue is therefore indispensable. The use of non-tissue equivalent material for this purpose may induce substantial bias and misleading conclusions (Schneider *et al* 1996, Wohlfahrt *et al* 2017a).

The conventional SECT-based approach has been validated in animal tissue in a view published studies (Schaffner and Pedroni 1998, Rietzel *et al* 2007, Zhang *et al* 2017). All of them performed ion transmission experiments involving rather long beam paths in tissue of the order of 10 cm. Schaffner & Pedroni, as well as Rietzel *et al.* tried to select beam paths of "maximal homogeneity" in the CT image and compared the predicted mean SPR to the reference SPR obtained from measured depth dose curves. Zhang *et al.* compared the results of a ray tracing algorithm based on a CT scan with a recorded 2D radiography. The experimental methods were of limited applicability for heterogeneous tissue, in particular for bone. For soft tissue, the precision and accuracy can be considered sufficient for the intended purpose, that is, to validate a single HLUT with an uncertainty on the percent level.

For a meaningful comparison of different SPR prediction methods (e.g., SECT vs. DECT), however, an accuracy of reference SPR at the per mil level is desirable, since differences are expected to be at or below the percent level. Such differences in SPR can translate to clinically relevant range deviations in patients (Wohlfahrt *et al* 2017d). A convenient technology to provide an SPR reference could be direct 3D SPR imaging via proton or ion CT (Penfold *et al* 2009, Poludniowski *et al* 2015). Unfortunately, such systems are still under development and currently not able to deliver sufficient SPR accuracy to serve as benchmark for x-ray CT (Johnson *et al* 2016). Currently, it is therefore indispensable to rely on transmission experiments without 3D reconstruction, which in turn poses severe constraints on the experimental setup. For example, the tissue has to be prepared in a volume between two plane-parallel boundaries (for entry and exit of the beam). Furthermore, the samples need to be homogeneous across at least the typical lateral profile of a clinical ion beam (5 – 10 mm) to avoid Bragg peak degradation and to allow for a coherent interpretation of the results on the SPR level.

In this work, we therefore introduce an experimental design for CT imaging and particle transmission measurements involving a dedicated 3D-printed sample container. Using this setup, we acquired reference SPRs for several animal tissues and compared them with SPR predictions based on both single- and dual-energy CT scans.

## 2 Materials and methods

### 2.1 Experiments

Using a 3D-printed sample container (section 2.1.1), three series of measurements with different groups of tissue samples were performed within one day, respectively. For each series, the samples were prepared in the morning (section 2.1.2), CT scans were acquired in the afternoon (section 2.1.3), and ion transmission measurements were performed in the evening and at night (section 2.1.4).

#### 2.1.1 Sample-container design

We designed and built several copies of a sample container using the 3D printer Objet30 Pro (Stratasys, Eden Prairie, Minnesota, USA). One container consists of 14 serially aligned chambers with a prismatic inner volume of 15 x 17.8 x 17.8 mm$^3$. Each chamber can be sealed individually with a lid to avoid interaction of the sample with its environment. The container is readily adapted to meet the requirements for optimal CT and ion-beam measurements (Figure 1).

#### 2.1.2 Preparation of tissue samples

Thirteen different porcine and bovine tissues were selected for measurement (Table 1) according to criteria such as their relative abundance in the body, homogeneity and general availability. Five cuboids were cut out of each investigated animal tissue and fit into five chambers belonging to the same sample container. The rigid bone tissue was cut using a pad saw, which was operated on minimal speed to avoid tissue burning. We were particularly careful to tightly confine the samples between the two plane-parallel boundary surfaces of the sample chamber that would be perpendicular to the ion-beam direction in our setup (Figure 1D). This was done to ensure a uniform water-equivalent thickness across at least the lateral beam profile (~5-10 mm). For this, a drain tube was used under light pressure on the soft tissue samples to remove the remaining air at the contact surfaces between the sample and the chamber walls. The thickness of the bone samples was adjusted by smoothing them down to appropriate size using abrasive paper. Small inclusions of air at other surfaces within the sample chamber (i.e., which would not be crossed by the ion beam) were accepted, as they do not influence the results.

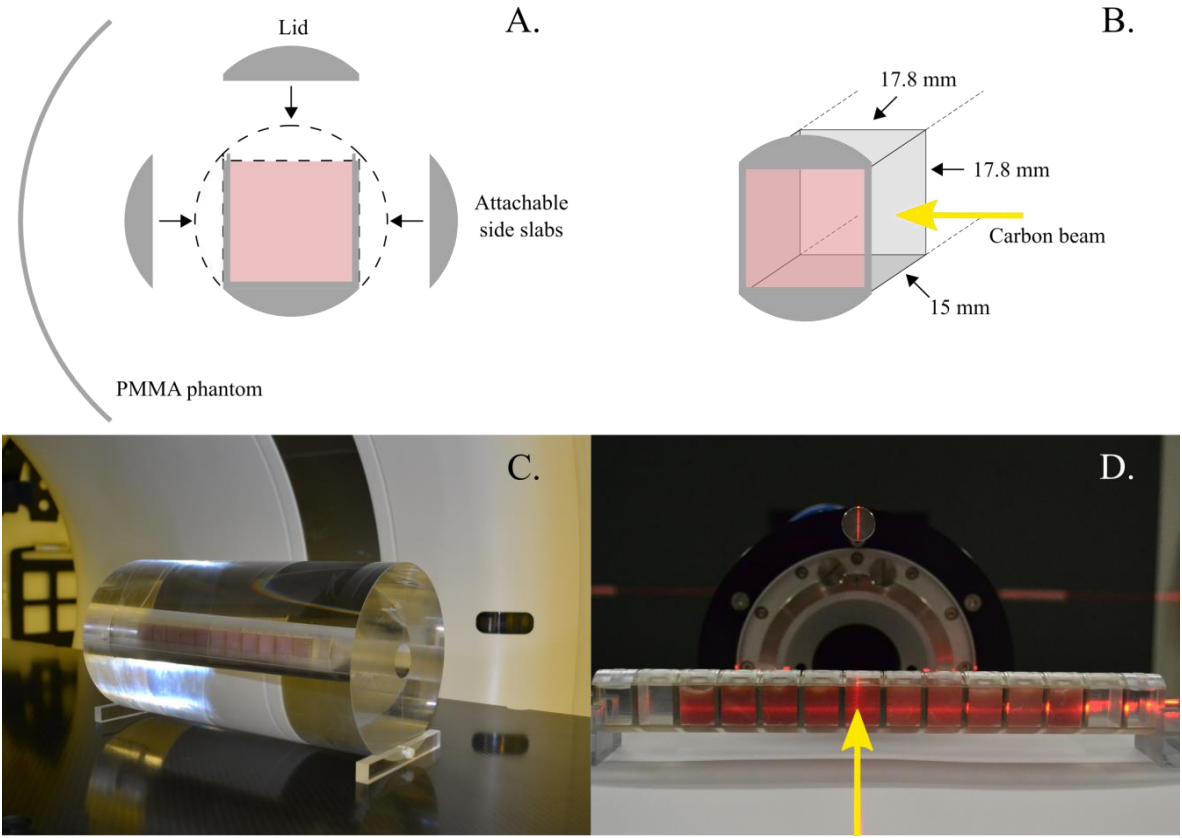

**Figure 1.** Configuration of the 3D-printed sample container (A, B) in the respective experimental setup (C, D). With a circular outer cross-sectional shape, the sample container fits tightly into the cylindrical bore of a PMMA phantom for CT measurements (A, C). For ion transmission measurements, the side-slabs are removed, in order to obtain two parallel boundary surfaces for entry and exit of the beam (B, D).

**Table 1**. Specifications of the investigated animal tissue samples. For each tissue class, five individual samples were prepared. The day of measurement is also indicated.

| Reference ID | Specification | Day |
|---|---|---|
| Adipose 1 | porcine back fat, smoked | 1 |
| Adipose 2 | porcine belly fat, fresh | 2 |
| Adipose 3 | bovine bone marrow, hung for a few weeks | 1 |
| Muscle 1 | porcine loin, fresh | 1 |
| Muscle 2 | porcine fillet, fresh | 1 |
| Muscle 3 | porcine thick flank, fresh | 1 |
| Liver | porcine liver, fresh | 2 |
| Kidney | porcine kidney, fresh | 3 |
| Brain | porcine brain, fresh | 3 |
| Heart | porcine heart, fresh | 3 |
| Blood | porcine blood, fresh | 3 |
| Lung | porcine lung, fresh | 2 |
| Bone | bovine cortical bone, sterilized | 3 |

In each container, a maximum of ten chambers were filled with two different types of animal tissue samples (five chambers each). Another two chambers were filled with de-ionized water and two were left empty to serve as reference for the calculation of SPR from measured depth-dose curves (section 2.3). The position of the chambers filled with these reference samples (air, water) was varied within the different containers, to check for systematic effects in the ion-range measurement.

After preparation and between experiments, the samples were stored cool (not frozen) without perceptible signs of degradation. To avoid temperature gradients in the samples during measurements, they were allowed to adapt to room temperature about one hour before data acquisition.

### 2.1.3 CT measurement

The samples were imaged with a Somatom Definition Flash DECT scanner (Siemens Healthineers, Forchheim, Germany). Two 3D-printed components were attached laterally to the sample container in order to complete a cylindrical form with a diameter of 2.8 cm (Figure 1A). The container could then be slid tightly into the central bore of a hollow cylindrical PMMA phantom with inner and outer diameters of 2.8 and 16 cm (Figure 1C). The entire imaged object thus resembles typical human dimensions in the transverse plane (e.g., head case) to enable stable and accurate beam hardening correction in the image reconstruction. With a length of 30 cm, the phantom was large enough to encase one sample container (~25 cm in length) at a time. It was positioned on the examination table such that the samples were aligned on the central axis in a reproducible position within the scanner.

SECT and DECT scans were acquired at tube potentials of 120 kVp and 80/140(Sn) kVp, respectively. The tube current time product was adjusted to 390 mAs for SECT and 701/351 mAs for DECT, to yield the same image dose in both scan modes ($CTDIvol_{16cm}$ = 59.7 mGy). Images were reconstructed with the Q34s\5 iterative reconstruction kernel using SAFIRE of maximal strength on a cubic grid of 0.6 mm edge length.

For use in the stoichiometric calibration, the tissue surrogates from a Gammex 467 phantom (Sun Nuclear Corporation, Melbourne, USA) were imaged employing the same experimental setup and CT scan protocol. The diameter of the sample container had been designed to match the diameter of the cylindrical Gammex inserts (2.8 cm), such that the same PMMA phantom could be used for both.

### 2.1.4 Ion transmission measurement

Ion transmission measurements were performed at the Ion Beam Therapy Center (HIT) in Heidelberg, Germany, using a carbon pencil beam with a kinetic energy of 200.28 MeV/u (corresponding to a range of 8.7 cm in water) and nominal focal spot size of 5.1 mm full width at half maximum (FWHM). The sample containers were placed on a conveyor orthogonal to the beam, which could be controlled from outside the room to speed up the experiment. The vertical position of the conveyor was adjusted such that the beam would pass shortly below the vertical center of the sample chamber. This was done to avoid potential inclusions of air originating from an insufficient filling level in the upper part of the chamber. The horizontal position was then adjusted to the center of each chamber in a reproducible way before the respective measurement. The conveyor movement was controlled via the laser positioning system and a video camera. For each measurement position, a depth-dose curve was recorded in a water absorber of variable thickness (PeakFinder, PTW, Freiburg, Germany) with a step size of 0.1 mm. With the optimized setup and workflow, the measurement time was about two minutes per sample.

### 2.2 SPR prediction and evaluation

#### 2.2.1 SECT approach

The 120 kVp SECT images were transformed into SPR datasets by voxel-wise application of a HLUT created by stoichiometric calibration (Schneider *et al* 1996). For the calibration, the mean measured CT number of each Gammex insert was evaluated in a cylindrical volume of interest. A regression on a CT-number-model according to Schneider et al. 1996 was performed based on the measured CT numbers and reference data on elemental composition and mass density (supplement, Table S1). CT numbers and SPR values were then calculated for a number of reference human tissues with tabulated elemental composition (Woodard and White 1986, White *et al* 1987). The CT numbers were derived from the calibrated stoichiometric model. The SPR was obtained from the Bethe formula using the Bragg additivity rule (Bragg and Kleeman 1905) for the superposition of I-values in compounds (Seltzer and Berger 1982). A kinetic energy of 200 MeV/u was chosen to match the initial beam energy used in the carbon transmission measurements. The reference tissues were interpolated by six line segments to obtain a HLUT (Figure 2). The look-up table in its numeric form is provided in supplement Table S2.

#### 2.2.2 DECT approach

The investigated method of DECT-based SPR prediction combines accurate and robust electron-density determination (Möhler *et al* 2017) with an empirical look-up table of the stopping number from the photon absorption cross section (Möhler *et al* 2016). The latter was calibrated based on calculations for the same reference tissues as used in the stoichiometric calibration in the previous section. The monoenergetic photon absorption cross section at 60 keV was determined by superposing elemental cross sections from the NIST Photon cross sections database (Berger *et al* 1998) according to elemental composition. The stopping number at 200 MeV/u was derived from the Bethe formula. Analogous to the HLUT case, the reference tissues were interpolated to form a look-up table (Figure 2, Table S3).

The practical implementation of the DECT approach used in this study is illustrated in Figure 3. From the 80/140(Sn) kVp DECT scan, an electron-density ("Rho") and a pseudo-monoenergetic image at 60 keV ("MonoCT") were calculated using the applications syngo.CT DE Rho/Z (Hünemohr *et al* 2014) and syngo.CT DE Monoenergetic Plus (Grant *et al* 2014) of the clinical image post-processing software syngo.via (version VB10B, Siemens Healthineers). Subsequently, these images were further processed by voxel-wise elementary arithmetic. First, the cross section was determined dividing MonoCT by Rho. For Rho < 0.3 (mainly air), the cross section was set to one to avoid instability in the division. The stopping number was then derived from the cross section via the look-up table described above and multiplied with the electron-density to obtain SPR in the final step.

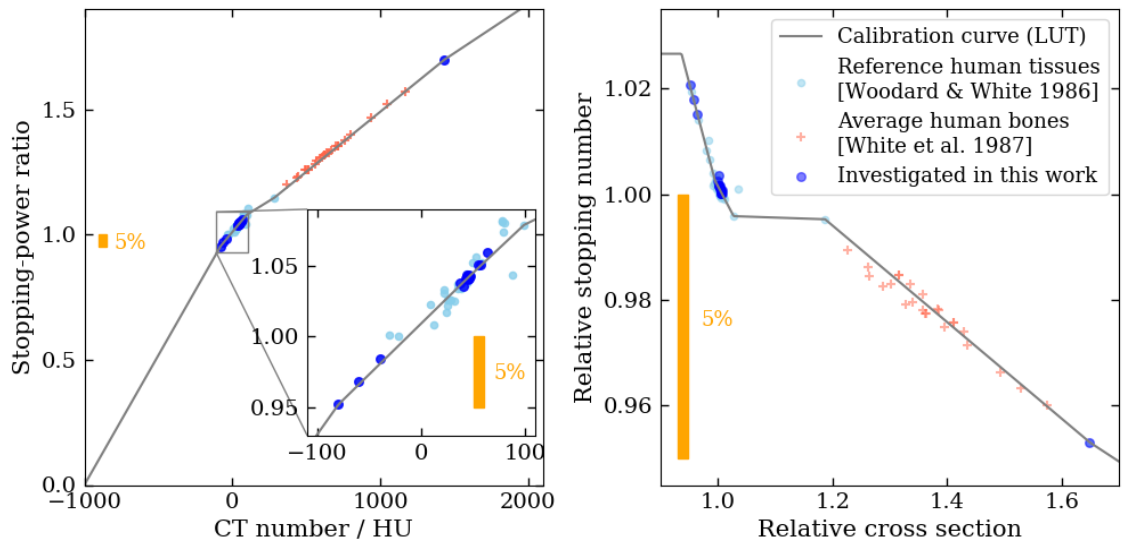

**Figure 2.** Calibration of the empirical component of the investigated SECT (left) and DECT (right) method. The corresponding look-up tables (LUTs) are provided in supplement (Tables S2, S3). The tissue classes investigated in this work (darker blue) are well fitted by the calibration curve without bias to ensure a fair comparison (see discussion). The orange 5% scale bars illustrate the different empirical influence on the respective method.

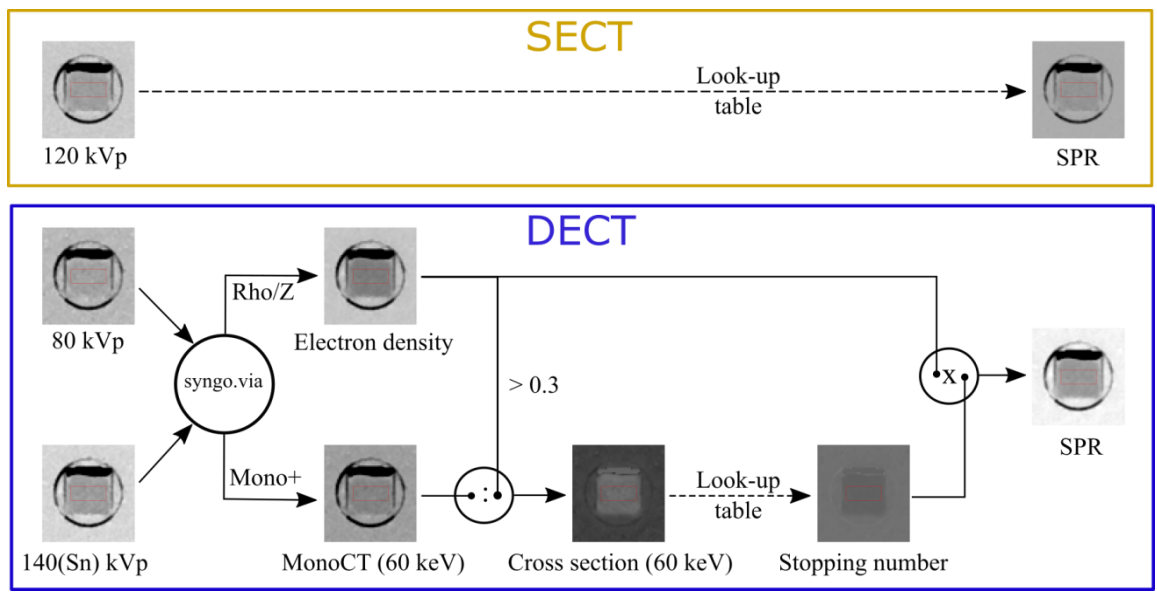

**Figure 3.** Processing of the CT images for SPR prediction. The exemplarily shown images (axial slices) are from one of the Muscle 1 samples. The gray values represent Hounsfield(-like) units (transformation $x' = (x - 1) \cdot 1000$). All images share a common grayscale window (C = 35 HU, W = 300 HU).

### 2.2.3 Image evaluation in volumes of interest

The mean and standard deviation of the respective voxel values (CT number, SPR, intermediate steps) were evaluated for each sample within a segmentation of cylindrical shape with a diameter of 6.5 mm (corresponding to about $3\sigma$ of a Gaussian beam spot with FWHM = 5.1 mm) and a total volume of approximately 400 mm$^3$ (1840 voxels). The segmentations were placed in conformity with the ion-beam track during range measurements, as visualized by the laser system.

### 2.3 Experimental determination of reference SPR

From a cubic spline interpolation of each recorded depth-dose curve, an estimator for the mean particle range, $r$, was extracted at the position of the distal fall-off where the relative dose is reduced to 80% of the maximum. The peak width, $w$, at 80% relative dose was accordingly determined to serve as measure for range degradation. The stopping-power ratio, $\hat{S}_{\text{ref}}$, of the sample was then obtained via a ratio of measured range differences,

$$\hat{S}_{\text{ref}} = \frac{r_a - r_s}{r_a - r_w}, \qquad (1)$$

with indices s, a and w for sample, air and water, respectively. Propagation of uncertainty in equation 1 yields the relative standard uncertainty (JCGM 2008)

$$u(\hat{S}_{\text{ref}})/\hat{S}_{\text{ref}} = \frac{[(r_a - r_w)^2 u(r_s)^2 + (r_a - r_s)^2 u(r_w)^2 + (r_s - r_w)^2 u(r_a)^2]^{1/2}}{(r_a - r_s)(r_a - r_w)}. \qquad (2)$$

It is reasonable to assume the same uncertainty in a single range measurement, $u(r)$, independent of the material in the chamber. In case of $N_x$ repeated measurements of a chamber with the same filling $x$, the uncertainty decreases according to $u(r_x) = u(r)/\sqrt{N_x}$. For a material similar to water ($r_s \approx r_w$) and a nominal sample thickness of $d \approx r_a - r_w$, equation 2 then simplifies to

$$u(\hat{S}_{\text{ref}})/\hat{S}_{\text{ref}} = \frac{u(r)}{d}\left(\frac{1}{N_s} + \frac{1}{N_w}\right)^{\frac{1}{2}}. \qquad (3)$$

## 3 Results

### 3.1 Uncertainty of reference SPR

For the air- and water-filled reference chambers, no systematic effect of the position within a sample container or between different sample containers was observed in measured range. Comparing the three days of measurements, however, small systematic shifts in absolute range (50 μm) between day 1 and 2 and considerable shifts (4 mm) between day 1 and 3 were observed for both the air and water reference samples (Table 2, see discussion). Therefore, their daily means were used in the calculation of SPR from equation 1. The standard deviations of measured ranges for all air- and water-filled chambers within one day was below 0.01 mm. A conservative estimate of $u(r) = 0.01$ mm was therefore used as single range uncertainty in the determination of SPR uncertainty via equation 2. The resulting uncertainty, as reported in the last column of Table 2, is well below 0.1% for all investigated samples except lung (0.13%) due to its low density.

*3.2 Accuracy of CT-based SPR prediction*

Deviations of SPR prediction with SECT and DECT (Table 3) from the measured reference SPR (Table 2) are reported together with corresponding significance tests in Table 4 and displayed in Figure 4.

For SECT, the deviations are significant for all individual tissue types except brain and bone. Consistently, the mean absolute prediction error for the combined tissue classes is $(1.27 \pm 0.12)\%$. Furthermore, the mean signed prediction error of $(-0.84 \pm 0.12)\%$ reveals an overall negative bias. The largest underestimations of -2.1% to -2.7% are obtained for the adipose tissues.

In contrast, the DECT predictions are consistent with the reference for all tissue types except Adipose 3, for which a significant deviation of $(0.19 \pm 0.11)\%$ is found, which is also the largest deviation of all tissue classes. Consequently, the mean absolute prediction error of $(0.10 \pm 0.15)\%$ and the mean signed prediction error of $(-0.02 \pm 0.15)\%$ are consistent with zero deviation from the reference SPR to a level of 0.15%.

For brain, lung and bone no significant differences between SECT and DECT are observed (Table 4, last column). For all other tissues the DECT approach performs significantly better.

The uncertainty in the SPR comparison, as estimated from the standard deviation in the prediction errors of the five samples of each tissue class (Table 4, shaded boxes in Figure 4) was between 0.1% and 0.4% for most tissues. Besides the contribution of 0.1% from the reference SPR (section 3.1), a part of this uncertainty can be ascribed to small deviations from complete homogeneity of the sample. This can lead to deviations in the mean SPR as slightly different parts of the sample might be probed by the ion beam and evaluated in the CT images. The intrinsic heterogeneity of the lung and bone samples leads to an uncertainty in SPR comparison one order of magnitude higher (~1-5%) than for the other, soft-tissue samples. The heterogeneity is reflected in the ion transmission measurement by considerably broadened Bragg peaks of about 1.5 mm for bone and 3 mm for lung compared to the water reference at 0.9 mm (Table 2). In contrast, the Bragg peak width increased only marginally for the other tissue samples (maximum relative increase of 7% in the case of liver), confirming a high grade of homogeneity in this case. The heterogeneity of lung and bone is also reflected by large standard deviations within the evaluated image regions (~100 HU and ~35-55 HU), clearly above the noise level (~5 HU) (Table S4).

**Table 2.** Measured particle range ($r$), Bragg peak width at 80% relative dose ($w$), derived reference SPR ($\hat{S}_{\text{ref}}$) and relative standard uncertainty ($u(\hat{S}_{\text{ref}})/\hat{S}_{\text{ref}}$, calculated via equation 2) of the investigated samples. For each tissue type, the mean ± standard deviation of $N$ measured samples is given.

| Reference ID ($N$) | $r$ [mm] | $w$ [mm] | $\hat{S}_{\text{ref}}$ | $u(\hat{S}_{\text{ref}})/\hat{S}_{\text{ref}}$ (%) |
|---|---|---|---|---|
| Air day 1 (6) | 83.991 ± 0.007 | 0.882 ± 0.008 | 0.0000 ± 0.0004 | - |
| Air day 2 (8) | 83.942 ± 0.008 | 0.884 ± 0.008 | 0.0000 ± 0.0005 | - |
| Air day 3 (7) | 79.835 ± 0.010 | 0.878 ± 0.005 | 0.0000 ± 0.0006 | - |
| Water day 1 (4) | 66.266 ± 0.004 | 0.884 ± 0.007 | 1.0000 ± 0.0002 | 0.06 |
| Water day 2 (7) | 66.210 ± 0.008 | 0.882 ± 0.008 | 1.0000 ± 0.0005 | 0.06 |
| Water day 3 (7) | 62.119 ± 0.010 | 0.873 ± 0.009 | 1.0000 ± 0.0006 | 0.06 |
| Adipose 1 (4) | 66.881 ± 0.042 | 0.902 ± 0.016 | 0.9653 ± 0.0024 | 0.06 |
| Adipose 2 (5) | 66.572 ± 0.049 | 0.904 ± 0.014 | 0.9796 ± 0.0028 | 0.06 |
| Adipose 3 (5) | 67.063 ± 0.128 | 0.897 ± 0.007 | 0.9550 ± 0.0072 | 0.07 |
| Muscle 1 (5) | 65.103 ± 0.010 | 0.896 ± 0.006 | 1.0656 ± 0.0006 | 0.06 |
| Muscle 2 (5) | 65.150 ± 0.034 | 0.903 ± 0.012 | 1.0629 ± 0.0019 | 0.06 |
| Muscle 3 (5) | 65.142 ± 0.010 | 0.909 ± 0.010 | 1.0634 ± 0.0006 | 0.06 |
| Liver (5) | 64.912 ± 0.016 | 0.945 ± 0.017 | 1.0732 ± 0.0009 | 0.06 |
| Kidney (5) | 61.320 ± 0.020 | 0.883 ± 0.010 | 1.0451 ± 0.0011 | 0.06 |
| Brain (5) | 61.494 ± 0.020 | 0.887 ± 0.022 | 1.0353 ± 0.0011 | 0.06 |
| Heart (5) | 61.222 ± 0.015 | 0.881 ± 0.006 | 1.0506 ± 0.0008 | 0.06 |
| Blood (5) | 61.065 ± 0.006 | 0.876 ± 0.005 | 1.0595 ± 0.0004 | 0.06 |
| Lung (5) | 75.688 ± 0.912 | 3.338 ± 0.517 | 0.4655 ± 0.0514 | 0.13 |
| Bone (5) | 49.322 ± 0.602 | 1.478 ± 0.230 | 1.7224 ± 0.0340 | 0.04 |

**Table 3.** Mean CT numbers and predicted SPRs in the evaluated image regions for SECT and DECT. For each tissue type, the mean ± standard deviation of $N$ measured samples is given.

|  | SECT | | DECT | | |
|---|---|---|---|---|---|
| Reference ID ($N$) | 120 kVp | $\hat{S}_{SE}$ | 80 kVp | 140(Sn) kVp | $\hat{S}_{DE}$ |
| Air day 1 (6) | -978.0 ± 2.4 | 0.0220 ± 0.0024 | -1003.6 ± 0.6 | -1001.9 ± 0.3 | -0.0037 ± 0.0004 |
| Air day 2 (8) | -977.2 ± 0.7 | 0.0229 ± 0.0007 | -1003.8 ± 0.2 | -1001.3 ± 0.5 | -0.0035 ± 0.0004 |
| Air day 3 (7) | -975.8 ± 1.0 | 0.0244 ± 0.0011 | -1005.0 ± 1.2 | -1002.7 ± 1.0 | -0.0049 ± 0.0012 |
| Water day 1 (4) | -1.2 ± 0.7 | 1.0080 ± 0.0005 | 4.5 ± 2.8 | 3.7 ± 2.8 | 1.0039 ± 0.0031 |
| Water day 2 (7) | -0.8 ± 1.3 | 1.0083 ± 0.0009 | 5.6 ± 2.6 | 6.1 ± 1.4 | 1.0073 ± 0.0010 |
| Water day 3 (7) | 0.7 ± 0.9 | 1.0094 ± 0.0007 | 2.6 ± 1.6 | 3.1 ± 2.0 | 1.0042 ± 0.0024 |
| Adipose 1 (4) | -86.4 ± 2.6 | 0.9446 ± 0.0025 | -111.1 ± 5.3 | -73.4 ± 4.2 | 0.9639 ± 0.0030 |
| Adipose 2 (5) | -76.1 ± 7.6 | 0.9534 ± 0.0062 | -97.8 ± 8.4 | -57.9 ± 6.3 | 0.9811 ± 0.0050 |
| Adipose 3 (5) | -97.9 ± 7.1 | 0.9330 ± 0.0072 | -125.4 ± 5.4 | -83.9 ± 6.9 | 0.9568 ± 0.0082 |
| Muscle 1 (5) | 67.6 ± 1.5 | 1.0567 ± 0.0010 | 73.8 ± 4.3 | 68.4 ± 3.9 | 1.0653 ± 0.0036 |
| Muscle 2 (5) | 65.5 ± 2.5 | 1.0553 ± 0.0017 | 70.1 ± 2.2 | 64.6 ± 1.8 | 1.0613 ± 0.0018 |
| Muscle 3 (5) | 65.7 ± 1.6 | 1.0554 ± 0.0011 | 71.6 ± 4.1 | 66.1 ± 3.1 | 1.0629 ± 0.0026 |
| Liver (5) | 77.0 ± 1.8 | 1.0632 ± 0.0012 | 83.0 ± 2.2 | 77.9 ± 2.1 | 1.0750 ± 0.0021 |
| Kidney (5) | 48.1 ± 1.6 | 1.0433 ± 0.0011 | 52.9 ± 1.1 | 48.2 ± 1.0 | 1.0454 ± 0.0010 |
| Brain (5) | 38.9 ± 1.3 | 1.0368 ± 0.0010 | 38.4 ± 1.9 | 36.5 ± 1.2 | 1.0357 ± 0.0015 |
| Heart (5) | 53.1 ± 1.6 | 1.0468 ± 0.0011 | 57.4 ± 1.2 | 52.5 ± 1.1 | 1.0496 ± 0.0011 |
| Blood (5) | 67.2 ± 1.5 | 1.0565 ± 0.0010 | 75.3 ± 2.7 | 65.5 ± 1.8 | 1.0590 ± 0.0014 |
| Lung (5) | -516.5 ± 70.3 | 0.4998 ± 0.0727 | -531.1 ± 73.5 | -532.4 ± 73.0 | 0.4680 ± 0.0730 |
| Bone (5) | 1559.4 ± 122.7 | 1.7450 ± 0.0493 | 2239.1 ± 171.8 | 1257.9 ± 94.3 | 1.7203 ± 0.0511 |

**Table 4.** Deviation of SECT- and DECT-based SPR prediction (Table 3) from the measured reference (Table 2). For each tissue group, the mean ± standard deviation is provided. For the calculation of the mean (absolute) error, the tissue classes were weighted by their respective contribution to the body mass (ICRP 1975). The rightmost column shows results of a two-sided t-test on the deviation from zero of SECT ($p_{SE}$) and DECT ($p_{DE}$) prediction errors as well as their difference ($p_{SE-DE}$). ns = not significant, * $p < 0.05$, ** $p < 0.01$, *** $p < 0.001$.

| Reference ID | rel. weight | $\frac{\hat{s}_{SE}}{\hat{s}_{ref}} - 1$ (%) | $\frac{\hat{s}_{DE}}{\hat{s}_{ref}} - 1$ (%) | $p_{SE}$ / $p_{DE}$ / $p_{SE-DE}$ |
|---|---|---|---|---|
| Adipose 1 | 0.09 | -2.14 ± 0.09 | -0.14 ± 0.24 | *** / ns / *** |
| Adipose 2 | 0.09 | -2.67 ± 0.38 | 0.16 ± 0.38 | *** / ns / *** |
| Adipose 3 | 0.09 | -2.31 ± 0.15 | 0.19 ± 0.11 | *** / * / *** |
| Muscle 1 | 0.16 | -0.83 ± 0.09 | -0.02 ± 0.30 | *** / ns / *** |
| Muscle 2 | 0.16 | -0.71 ± 0.18 | -0.15 ± 0.14 | *** / ns / *** |
| Muscle 3 | 0.16 | -0.75 ± 0.11 | -0.05 ± 0.23 | *** / ns / *** |
| Liver | 0.03 | -0.93 ± 0.11 | 0.17 ± 0.21 | *** / ns / *** |
| Kidney | 0.01 | -0.17 ± 0.11 | 0.03 ± 0.08 | * / ns / * |
| Brain | 0.02 | 0.14 ± 0.16 | 0.05 ± 0.16 | ns / ns / ns |
| Heart | 0.01 | -0.37 ± 0.14 | -0.09 ± 0.11 | ** / ns / ** |
| Blood | 0.10 | -0.29 ± 0.08 | -0.04 ± 0.15 | ** / ns / * |
| Lung | 0.02 | 7.05 ± 4.41 | 0.15 ± 5.15 | * / ns / ns |
| Bone | 0.07 | 1.30 ± 1.17 | -0.14 ± 1.31 | ns / ns / ns |
| mean error | - | -0.84 ± 0.12 | -0.02 ± 0.15 | - |
| mean absolute error | - | 1.27 ± 0.12 | 0.10 ± 0.15 | - |

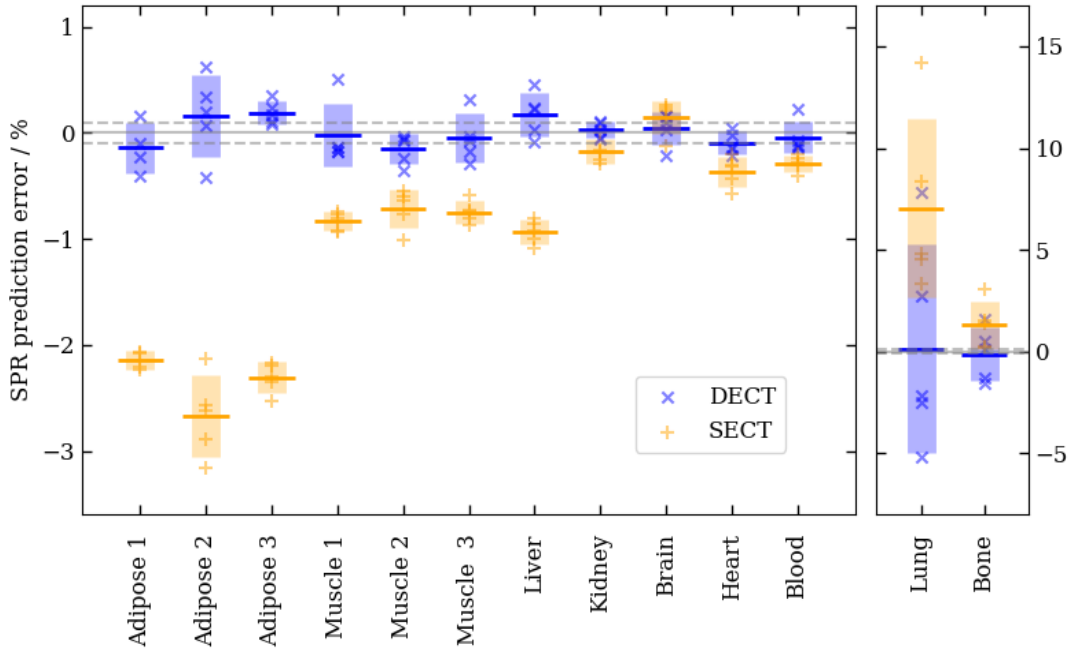

**Figure 4.** Deviations of SECT- and DECT-based SPR predictions from the measured reference. Positive (negative) deviation indicates over-(under-)estimation of the prediction. The shaded boxes correspond to plus/minus one standard deviation around the mean value (horizontal line). The gray dashed lines indicate the order of magnitude of the experimental uncertainty (0.1%).

## 4  Discussion

An experimental setup for consecutive CT scans and ion transmission measurements was introduced, enabling SPR determination with an uncertainty below 0.1%. The setup was used to evaluate CT-based SPR predictions from different methods in a large variety of animal tissues.

The small standard deviations of the range measurements across different chamber positions and containers confirm very high precision of the entire measurement chain, including the 3D printing of the sample containers, as well as the alignment of the setup. This also supports that sampling the depth dose curve in steps of 0.1 mm is sufficient to achieve reproducibility in measured range below 0.01 mm. This allows for a considerable increase in measurement speed and sample throughput compared to using the maximum sampling rate of 0.01 mm. In typical setups for ion transmission experiments (e.g., Schaffner & Pedroni 1998), a substantial source of systematic uncertainty is induced by the length of the beam path in the sample, which is hard to measure exactly. Using our proposed experimental design, the sample thickness is implicitly accounted for by the reference water and air samples of the same thickness (cf., equation 1). The constancy of this thickness is ensured to high precision by the 3D-printed sample container, so that the uncertainty component associated to sample thickness is practically removed.

The reference SPR uncertainty of below 0.1% is sufficient for the study purpose to discriminate different CT-based SPR prediction approaches. In principle, the measurement uncertainty could even be further reduced by either increasing the sample thickness or the number of measurements for the sample and the water reference, according to equation 3. An increase in sample thickness is not easily possible without compromising the CT setup (Figure 1). Furthermore, below the per mil level, other sources of error will become relevant, which are hard to correct for, such as temperature and air pressure effects or the energy dependence in SPR.

With the various investigated muscle and fat samples, selected inner organs, lung and cortical bone, the most important tissue families have been included, adding up to more than 80% of the human body mass (ICRP 1975). The full SPR range of soft tissues was covered (~0.95 – 1.07). Including tissues that are more heterogeneous by nature remains an experimental challenge, as demonstrated for the lung and bone samples in this study. In addition, bones are particularly difficult to prepare in accordance with the experimental requirements, as they are not only highly heterogeneous but also rigid in structure.

One specific static HLUT was evaluated, which might be sub-optimal for the measured tissues. On the other hand, a state-of-the-art stoichiometric calibration was applied. The fat and muscle reference points were even included in the definition of the HLUT. A fair comparison was also assured by the choice of similar base points in the definition of the SECT and DECT look-up tables (Tables S2, S3, Figure 2). Also, only one DECT approach was evaluated. However, the acquired ion-range and CT data (Tables 2, 3) can be used to test other image-based SPR algorithms (van Elmpt *et al* 2016) in a similar manner as done here.

The CT-based SPR predictions were obtained in an idealized geometry (all samples on the central axis in the CT scanner, only one sample per transverse plane, well controlled beam hardening situation).

Additional sources of uncertainty can therefore occur in a patient case. These additional effects of a realistic patient anatomy were analyzed in a complementary study using an anthropomorphic phantom (Wohlfahrt *et al* 2017c).

A positive bias in the water samples of around 5 HU was observed in the 80 kVp and 140(Sn) kVp images, whereas the 120 kVp images of the same water samples were centered at 0 HU. Most probably, this reflects the CT scanner calibration being optimized for 120 kVp. The bias in the CT numbers leads to mean SPR prediction errors for water of 0.4% for day 1 and 3 and 0.7% for day 2. However, these errors are still smaller than the corresponding SPR prediction error of around 0.8% for water in the case of SECT. Even though the original CT image is unbiased in this case, a positive error is introduced by the design of the HLUT (section 2.2.1). The SPR predictions of animal tissues seem not to be affected by a potential shift of the calibrated water point for DECT, possibly due to a subtraction effect in the calculation of electron density.

The direction and magnitude of the absolute range shifts between the different days were consistent for air and water reference samples and could also be reproduced by a control measurement without a sample container in the beam path. The small deviation of 50 μm between day 1 and 2 might be due to a day-to-day variation in beam energy or a varying amount of air traversed by the beam (the distance between the nozzle and the measurement device was not explicitly controlled). The larger deviation of 4 mm on day 3 is probably owing to the use of a second PeakFinder model, which was not calibrated for absolute range measurement. The differences in absolute range, however, do not affect SPR accuracy, since only daily means were used in equation 1.

In this study, a kinetic energy of 200 MeV/u was used for SPR prediction corresponding to the initial beam energy, that is, in accordance with the approximate energy at which the sample is traversed. For application in patients it is recommended to use an effective beam energy of 100 MeV/u in CT-based SPR prediction, which minimizes the error induced by not adapting SPR values to energy loss (Inaniwa and Kanematsu 2016). The corresponding look-up tables for 100 MeV/u are also included in supplement (Tables S2, S3). A maximum difference between the SPR predictions at 100 and 200 MeV/u of 0.15% (0.4%) can be expected for soft (bony) tissue.

The presented experimental setup was used here to compare SPR predictions from CT scans with a reference SPR derived from ion transmission measurements. Alternatively, the presented experimental method can be readily used to determine I-values by combining electron-density estimation from DECT with SPR measurement (Table 5).

**Table 5.** I-values calculated from the measured SPR and relative electron density (RED). The adipose and muscle tissues were joined into one single class, respectively. The I-value of water was set to 78 eV (ICRU 2014). Please note that the quoted standard deviations can reflect both the actual variability in the samples and the experimental uncertainty. The large standard deviations for lung and bone illustrate the experimental challenge for heterogeneous tissues. This is particularly relevant for I-value determination, since experimental uncertainties are exponentially enhanced by the inversion of the logarithmic term in the Bethe formula. An additional systematic uncertainty might arise from the current uncertainty in the water I-value (2 eV).

| Reference ID | SPR | RED | I-value [eV] |
| --- | --- | --- | --- |
| Adipose | 0.9667 ± 0.0117 | 0.9458 ± 0.0124 | 64.8 ± 1.5 |
| Muscle | 1.0640 ± 0.0016 | 1.0636 ± 0.0032 | 77.8 ± 1.5 |
| Liver | 1.0732 ± 0.0009 | 1.0754 ± 0.0020 | 79.3 ± 1.3 |
| Kidney | 1.0451 ± 0.0011 | 1.0457 ± 0.0010 | 78.4 ± 0.5 |
| Brain | 1.0353 ± 0.0011 | 1.0353 ± 0.0013 | 78.0 ± 1.1 |
| Heart | 1.0506 ± 0.0008 | 1.0500 ± 0.0011 | 77.6 ± 0.7 |
| Blood | 1.0595 ± 0.0004 | 1.0607 ± 0.0016 | 78.8 ± 1.1 |
| Lung | 0.4655 ± 0.0514 | 0.4679 ± 0.0730 | 83.5 ± 37.9 |
| Bone | 1.7224 ± 0.0340 | 1.8165 ± 0.0586 | 121.0 ± 14.8 |

# 5  Conclusion

An experimental setup was introduced to acquire ion-range and CT data on a representative selection of animal tissues. An uncertainty in the SPR reference of below 0.1% was achieved, which is unprecedented for tissue. The SECT predictions showed significant errors of around 1% on average, while the DECT predictions were consistent with the reference. The potential of DECT to decrease range uncertainty was clearly underlined by this study.

## Acknowledgments

This work was partially funded by the National Center for Radiation Research in Oncology (NCRO) and the Heidelberg Institute for Radiation Oncology (HIRO) within the project "translation of dual-energy CT into application in particle therapy".